\documentclass[preprintnumbers,article,amsmath,amssymb,floatfix,10pt,prd,onecolumn,
superscriptaddress,nofootinbib]{revtex4}
\usepackage{bbm}
\usepackage{amsfonts}
\usepackage{mathrsfs}
\usepackage{latexsym}

\usepackage{epsfig}
\usepackage{epstopdf}
\usepackage{graphicx}
\usepackage{amssymb}
\usepackage{amsmath}
\usepackage{dcolumn}
\usepackage{bm}
\usepackage{color}
\usepackage{comment}
\usepackage{xcolor}

\begin{document}

\title{\bf Traversable Wormhole Solutions in $f(R)$ Gravity Via Karmarkar Condition}
\author{M. Farasat Shamir}
\email{farasat.shamir@nu.edu.pk}\affiliation{National University of Computer and
Emerging Sciences,\\ Lahore Campus, Pakistan.}
\author{I. Fayyaz}
\email{iffat845@gmail.com}\affiliation{National University of Computer and
Emerging Sciences,\\ Lahore Campus, Pakistan.}

\begin{abstract}
Motivated by recent proposals of possible wormhole shape functions, we construct a wormhole shape function by employing the Karmarkar condition for static traversable wormhole geometry. The proposed shape function generates wormhole geometry that connects two asymptotically flat regions of spacetime and satisfies the required conditions. Further, we discuss the embedding diagram in three-dimensional Euclidean space to present the wormhole configurations.
The main feature of current study is to consider three well-known $f(R)$ gravity models, namely exponential gravity model, Starobinsky gravity Model and Tsujikawa $f(R)$ gravity model. Moreover, we investigate that our proposed shape function provides the wormhole solutions with less (or may be negligible) amount of exotic matter corresponding to the appropriate choice of $f(R)$ gravity models and suitable values of free parameters.
Interestingly, the solutions obtained for this shape function generate stable static spherically symmetric wormhole structure in the context of non-existence theorem in $f(R)$ gravity. This may lead to a better analytical representation of wormhole solutions in other modified gravities for the suggested shape function.

\end{abstract}

\maketitle

\date{\today}

\section{Literature Survey}


Wormholes are assumptive tube-like geometrical structures connecting two widely separated asymptotically flat distant universes, or two different asymptotically flat portions of the same universe and have no horizon. The concept of a tube-like or bridge-like structure connecting two spacetimes was firstly presented by Einstein and Rosen \cite{ein}. Further, they explored the exact solution that describe the geometrical structure of the bridge. The solution presented by Einstein and Rosen is linked with the work of Flamm \cite{fla}, who first time developed the isometric embedding of Schwarzschild solution. Ellis \cite{ell} introduced another term for wormholes which is known as $``Drainhole"$. Wheeler \cite{whe} named them as $``Geons"$ and predicted the shape of a wormhole which offers a twofold space. Static wormhole configurations have a constant throat radius while non-static wormhole configurations have a variable throat radius. Kar \cite{kar7} discussed the static wormhole and examined their properties along with examples. Kar and Sahdev \cite{kar72} explored evolving Lorentzian wormholes and discussed some results for exponential and Kaluza-Klein inflation. The locally anisotropic wormhole and flux tube like solutions have been studied using the technique of anholonomic frames \cite{vac}. Dzhunushaliev et al. \cite{Dzh} have investigated the linear stability analysis for wormhole configurations with and without electric and/or magnetic fields. Visser et al. \cite{Vis} have shown that the violation of null energy condition (NEC) can be restricted to an arbitrarily small region by an appropriate choice of the wormhole geometry. The static and spherically symmetric Lorentzian wormhole solutions are discussed within the frame work of general theory of relativity (GR) \cite{can7}.

Morris-Thorne \cite{mor1} were the first who gave the idea that human being can travel through wormhole tunnels. They investigated the static spherically symmetric wormholes by using the principles of GR and presented the fundamental theory for traversable wormholes. According to Morris et al. \cite{mor2} investigation, for the formation of wormhole structure, the presence of exotic matter plays an important role in the context of GR. Exotic matter is a form of dark energy (having an EoS with $\omega <-1/3$), produces a repulsive force. Recent observations have shown that the dark energy is the main reason for the accelerated expansion of the universe. A few candidates have been proposed in the literature to presenting dark energy, like, a positive cosmological constant, Chaplygin gas, the quintessence fields etc. GR has been the most successful theory of last century. One may simply needs to add a cosmological constant or some other exotic source to explain the accelerated expansion. However, there are models which may also explain dark energy by modifying gravity.
Recently, different researchers discussed wormholes geometry in different modified theories of gravity  \cite{nan,lob,ric,meh1,dzh,sha,meh2}. In $1970$, Buchdahl \cite{Buch} proposed one of the most well known and simplest modified theory of gravity i.e., $f(R)$ gravity which can be obtained by adopting a modification in the Einstein Hilbert action with an arbitrary function of the Ricci scalar $R$. Following in this way, several modifications of GR have been developed by considering different approaches. The geometry of dark energy model can be depicted by $f(R)$ gravity \cite{cap1,cap2}. Harko et al. \cite{har} discussed static spherically symmetric wormholes with non-exotic matter (respected energy conditions) in the framework of $f(R)$ gravity. Rahaman and collaborators \cite{rah,jam} developed new solutions for static wormholes in $f(R)$ gravity and acknowledged the existence of wormhole solutions that
do not require exotic matter. Bronnikov and Starobinsky \cite{bron7} discussed the very well known non-existence theorem for wormhole geometry and observed that no realistic wormhole can be formulated in scalar-tensor models for a positive scalar function. Further, Bronnikov et al. \cite{bro} showed that for $\frac{df}{dR}=F(R)<0$, the non-existence condition of wormhole could be violated in $f(R)$ theory of gravity. Bahamonde et al. \cite{bah} discussed the existence of wormholes in galactic halos. Moreover, they described a non-static wormhole geometry that asymptotically impending towards the Friedmann Lemaître Robertson Walker universe.

One interesting topic in traversable wormhole geometry is the study of wormhole shape function (WSF) $\epsilon(r)$ for asymptotically flat wormhole with their essential properties (see detail in Sec. $II$). Recently, many authors have discussed various ansatz shape functions to describe the wormhole structure. Godani and Samanta \cite{goda} explored asymptotically flat wormhole with WSF $\epsilon(r)=\frac{r_0 Log(r+1)}{Log(r_0 +1)}$. Jahromi and Moradpour \cite{jah} introduced a WSF $\epsilon(r)= a ~tanhr$. The WSF $\epsilon(r)=\alpha + \beta(r)$ is proposed by Cataldo and Liempi \cite{cata}. Samanta et al. \cite{sam} defined $\epsilon(r)= \frac{r}{e^{(r-r_0)}}$, known exponential WSF. Some authors \cite{jam2,shar2,sham3} used $\epsilon(r)=(r_0)^{n+1} r^{-n}$, for different values of $n$. Recently, Golchin et al. \cite{gol} have considered a special class in $f(R)$ theory of gravity ($F(R_0) = 0$ and $\frac{df}{dR}|_{(R_0)} = 0$, where $R_0$ is a fixed quantity) and construct a WSF.

Motivating from this, in this article, we calculate a WSF by employing the Karmarkar condition (KMc). Karmarkar \cite{karm} developed a mandatory condition for a static and spherically symmetric line element to be of class one. In recent years, different researchers have considered the KMc to discussed the configurations of spherically symmetric compact objects \cite{6abb, 6ful, 6kuh, 6bah, 6ged}. Kuhfittig \cite{kuhf} developed wormhole geometry using KMc and shown that the embedding theory may provide the basis for a complete wormhole solution. Recently, Fayyaz and Shamir \cite{fay1} constructed a WSF by applying KMc and analysed that it obeys all the required conditions and exhibit the presence of exotic matter in the context of GR. To the best of our knowledge, we first time apply this WSF in modified $f(R)$ theory of gravity. This paper is composed as follows. In Sec. \textbf{II}, we construct a WSF by employing KMc and discuss the embedding diagram for wormhole geometry into three-dimensional Euclidean space. In Sec. \textbf{III}, we discuss the formulation of $f(R)$ theory of gravity. Further, we consider three different well known viable $f(R)$ gravity models and examined the NEC and weak energy condition (WEC) via graphical representation. Conclusion is given in Sec. \textbf{IV}.

\section{The Geometry of Traversable Wormhole and Embedding Diagram}

In the present work, our main focus is to develop a WSF using the KMc that describes wormhole geometry. For this purpose, we assume static spherically symmetric spacetime defined as:

\begin{equation}\label{71}
ds^2=-e^{\chi (r)} dt^2 +e^{\eta (r)} dr^2 +r^2 d\theta^2 +r^2\sin^2\theta d\phi^2.
\end{equation}
The non-zero Riemann curvature coordinates corresponding to the above spacetime ($\ref{71}$), are given below:
\begin{equation}\label{72}\nonumber
R_{1414}=\frac{e^{\chi} (2 \chi'' + \chi'^2-\chi' \eta')}{4},~~~~~~~~~~~~~~~R_{1212}=\frac{r \eta'}{2},~~~~~~~~~~~~~~R_{1334}=R_{1224} Sin^2 \theta,
\end{equation}
\begin{equation}\label{73}\nonumber
R_{2323}=\frac{r^2 Sin^2 \theta (e^\eta -1)}{e^\eta},~~~~~~~~~~~~~~~~~~~~~~~~~~ R_{3434}=\frac{r Sin^2 \theta \eta' e^{\chi-\eta}}{2},~~~~~~~~~~~~ R_{1224}=0.
\end{equation}

These components satisfying the very well known Karmarkar relation
\begin{equation}\label{75}\nonumber
R_{1414}=\frac{R_{1212} R_{3434} + R_{1224} R_{1334}}{R_{2323}},
\end{equation}
with $R_{2323}\neq0$. The form of spacetime, satisfying the KMc is known as embedding class one. By substituting the non-zero components of
Riemann curvature in Karmarkar relation, we get the following differential equation:
\begin{equation}\label{76}\nonumber
\frac{\chi' \eta'}{1-e^{\eta}}=\chi' \eta'-2 \chi''-\chi'^2,
\end{equation}
where $e^{\eta}\neq1$. The solution of the above differential equation is given as
\begin{equation}\label{78}
e^{\eta}=1+ A e^{\chi} \chi'^2.
\end{equation}
Here, $A$ is an integrating constant. Now, to construct the WSF, we assume Morris-Thorne metric defined as:

\begin{figure}\center
\begin{tabular}{cccc}
\epsfig{file=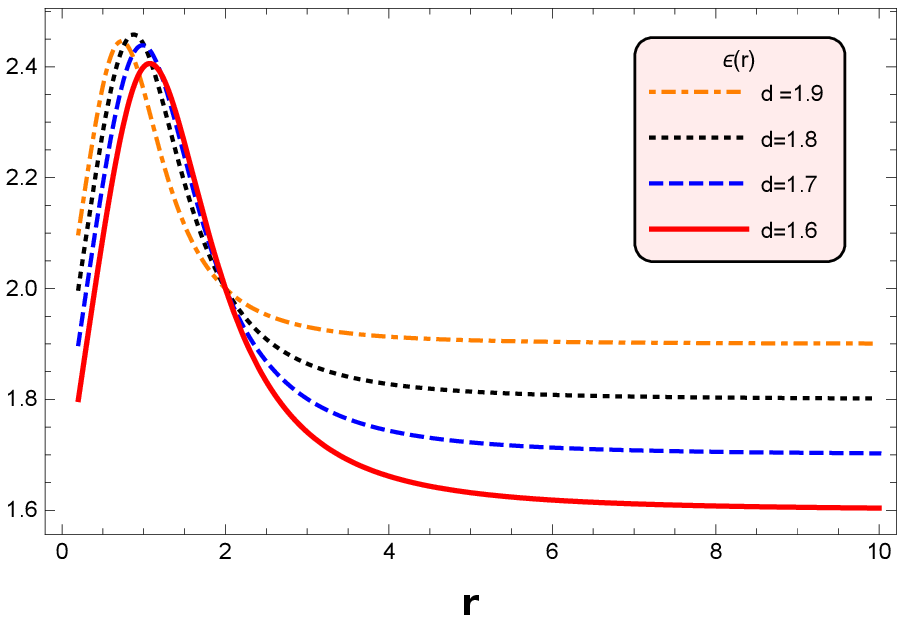,width=0.28\linewidth} &
\epsfig{file=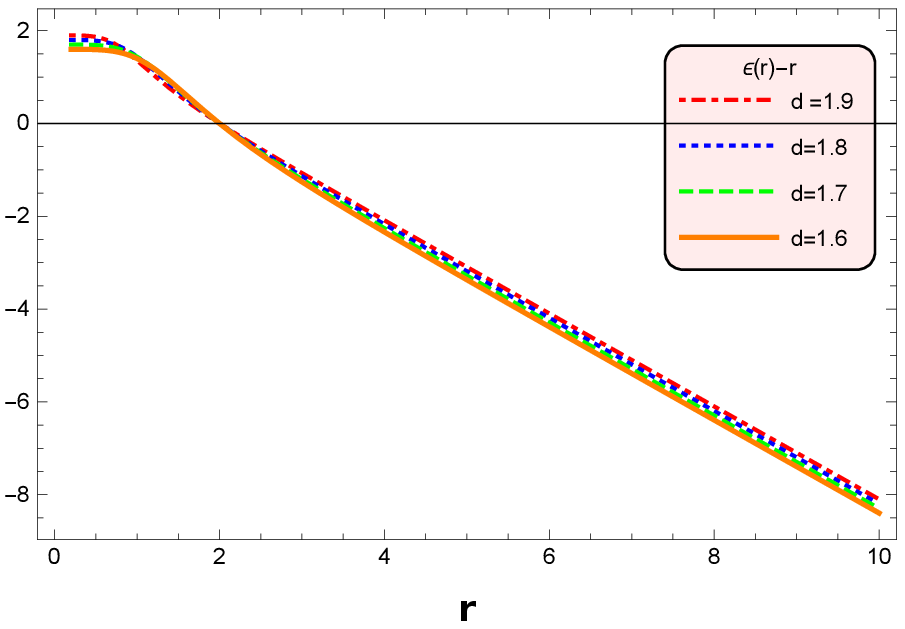,width=0.28\linewidth} &
\epsfig{file=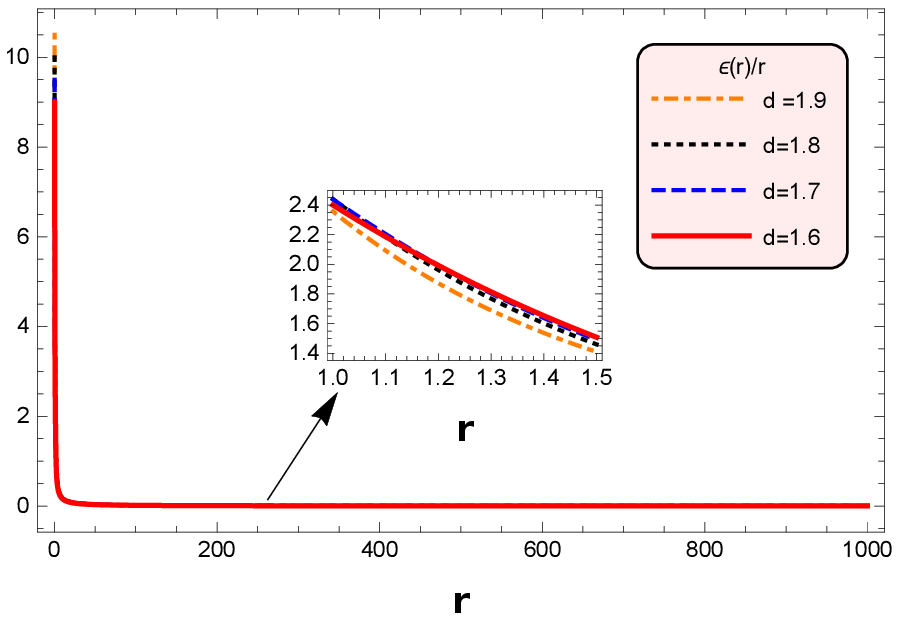,width=0.28\linewidth} \\
\end{tabular}
\caption{Evaluation of wormhole shape function $\epsilon(r)$ (Ist plot) and its required conditions $\epsilon(r)-r=0$ at $r =r_0$ (2nd plot), $\frac{\epsilon(r)}{r} \rightarrow 0$ for $r\rightarrow \infty$ (3rd plot) for $r_0=2$ and $C= 1.9$ to present the realistic wormhole geometry.}\center
\label{Fig:7a1}
\end{figure}

\begin{figure}\center
\begin{tabular}{cccc}
\epsfig{file=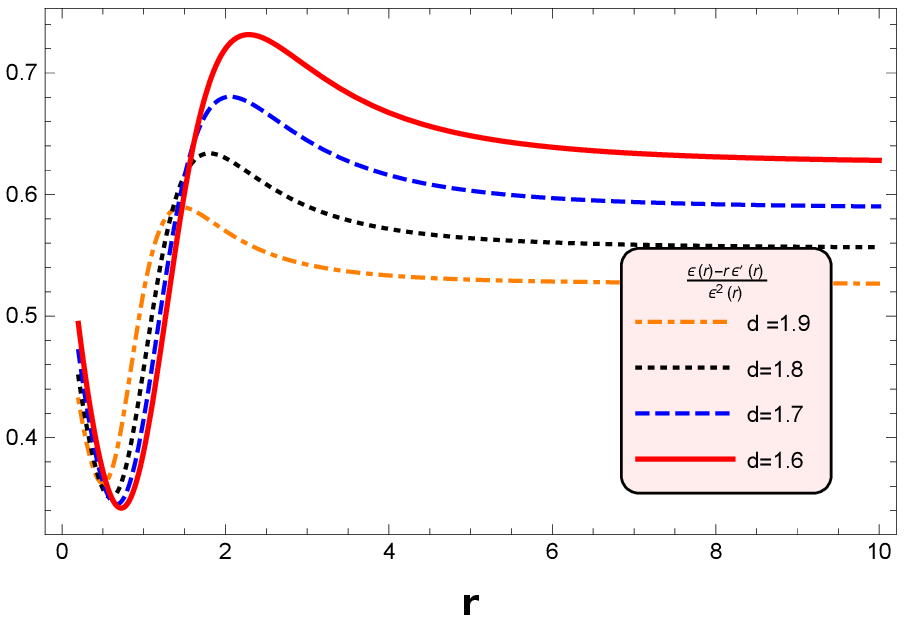,width=0.28\linewidth} &
\epsfig{file=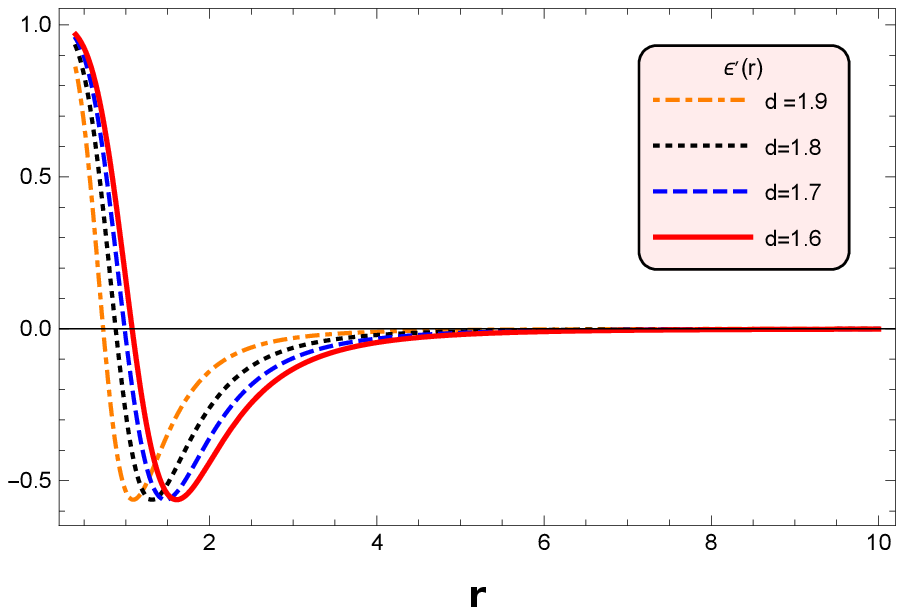,width=0.28\linewidth} \\
\end{tabular}
\caption{Condition $\frac{\epsilon(r)-r \epsilon'(r)}{\epsilon'(r)}>0$ (Ist plot) and $\epsilon'(r)<1$ (2nd plot),  for $r_0=2$ and $C= 1.9$.}\center
\label{Fig:7a2}
\end{figure}

\begin{equation}\label{79}
ds^2=-e^{\chi(r)} dt^2 + \frac{1}{1-\frac{\epsilon(r)}{r}} dr^2 +r^2 d\theta^2 +r^2\sin^2\theta d\phi^2.
\end{equation}
The metric co-efficient $\chi(r)$ is known as redshift function such that $\chi(r)\rightarrow 0$ as $r \rightarrow \infty$. Here, we assume the redshift function $\chi (r)$, defined as \cite{anch, sham3}
\begin{equation}\label{710}
\chi (r)=\frac{-2 \xi}{r},
\end{equation}
where $\xi$ is an arbitrary constant. By comparing the Eqns. ($\ref{71}$) and ($\ref{79}$), we obtain
\begin{equation}\label{711}
\eta(r)= Log [\frac{r}{r-\epsilon(r)}].
\end{equation}
Here, $\epsilon(r)$ is a WSF. Now, by using Eqns. ($\ref{78}$), ($\ref{710}$) and ($\ref{711}$), we calculate the following WSF
\begin{equation}\label{712}
\epsilon(r)=r- \frac{r^5}{r^4 + 4 \xi^2 A e^{\frac{-2 \xi}{r}}}.
\end{equation}
According to Morris and Thorne \cite{mor1}, to get a traversable wormhole solution, WSF should satisfy the following essential properties:\\
\begin{enumerate}

\item{$\epsilon(r)-r=0$ at $r=r_0$},

\item {The condition $\frac{\epsilon(r)-r \epsilon'(r)}{\epsilon'(r)}>0$ must be fulfilled at $r=r_0$},

\item{$\epsilon'(r)<1$},

\item{$\frac{\epsilon(r)}{r} \rightarrow 0$ at $r\rightarrow\infty$},\\
\end{enumerate}
where $r_0$ is known as a wormhole throat radius and $r$ is the radial coordinate such that $r_0 \leq r \leq \infty$. When we evaluate Eqn. ($\ref{712}$) at the throat i.e., $\epsilon (r_0)-r_0=0$, we get a trivial solution $r_0 =0$. To handle this problem we add a free parameter $``C"$ in Eqn. ($\ref{712}$). Now, Eqn ($\ref{712}$) takes the form $\epsilon(r)=r- \frac{r^5}{r^4 + 4 \xi^2 A e^{\frac{-2 \xi}{r}}}+C$. Condition ($1$) provides $A= \frac{r_{0}^{4}(r_0-C)}{4 \xi^2 e^{\frac{-2\xi^2}{r_0}}}$. After substituting the value of $A$ in Eqn. ($\ref{712}$), one can find the expression for WSF $\epsilon (r)$ as
\begin{equation}\label{713}
\epsilon(r)=r- \frac{r^5}{r^4 + r_{0}^4 (r_0 -C)}+C,~~~~~~~~~~~~~0<C<r_0.
\end{equation}
Conditions ($2$) and ($3$) also satisfy for the given range of $C$. Applying condition ($4$) on Eqn. ($\ref{713}$), we found that
\begin{equation}\label{614}
\lim_{r \rightarrow \infty} \frac{\epsilon(r)}{r}=0.
\end{equation}
\begin{figure}\center
\begin{tabular}{cccc}
\epsfig{file=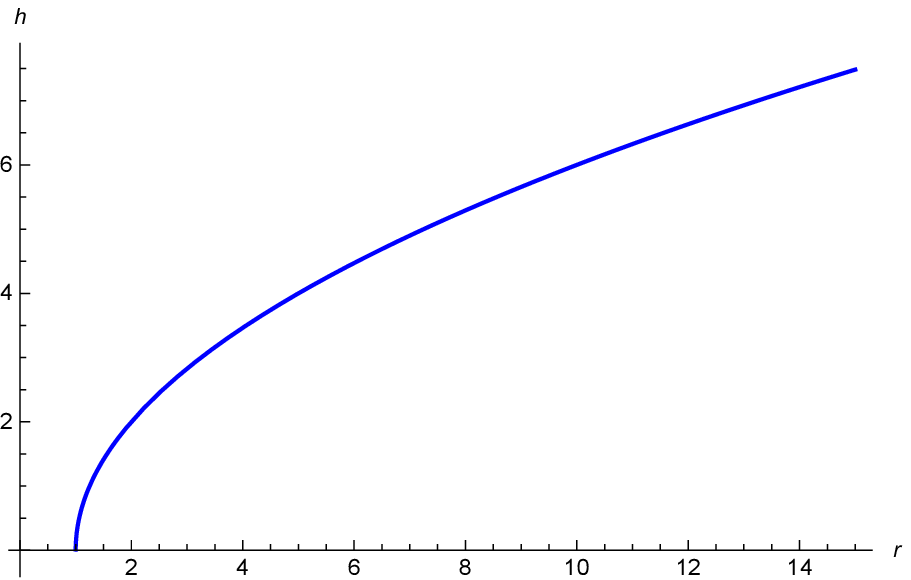,width=0.3\linewidth} &
\epsfig{file=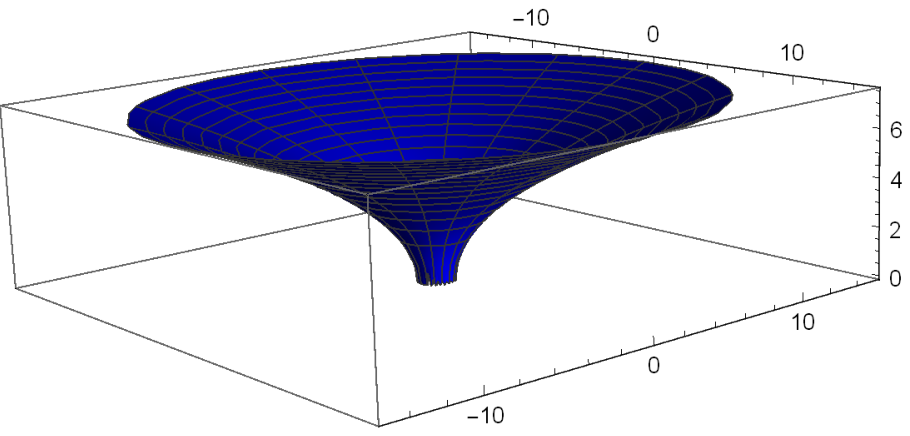,width=0.35\linewidth} &
\end{tabular}
\caption{Plot of embedding diagram for upper universe $h(r)>0$ with respect to radial coordinate with slice $t = const$ and
$\theta = \pi/2$. For a full visualization of the wormhole surface, took a $2\pi$ rotation around the $h-$axis.}\center
\label{Fig:7a3}
\end{figure}

\begin{figure}\center
\begin{tabular}{cccc}
\epsfig{file=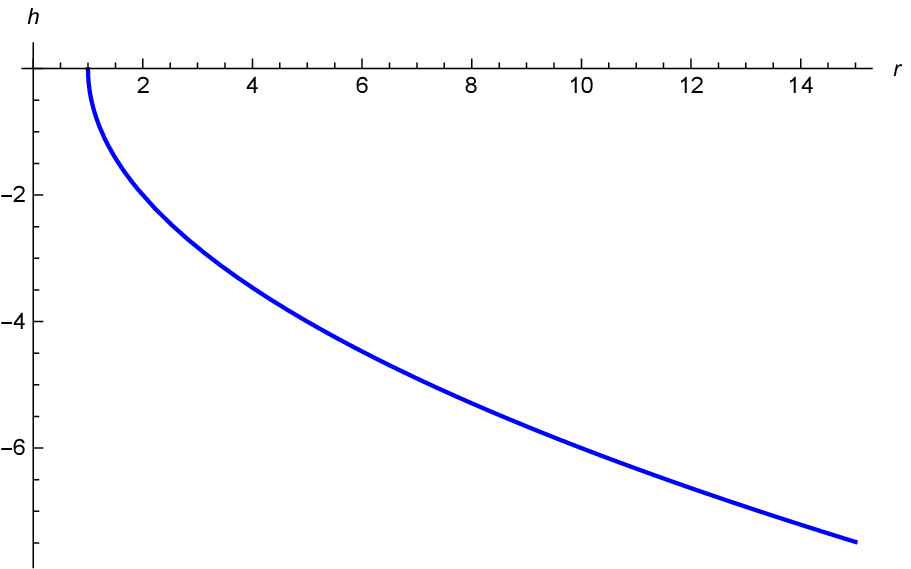,width=0.3\linewidth} &
\epsfig{file=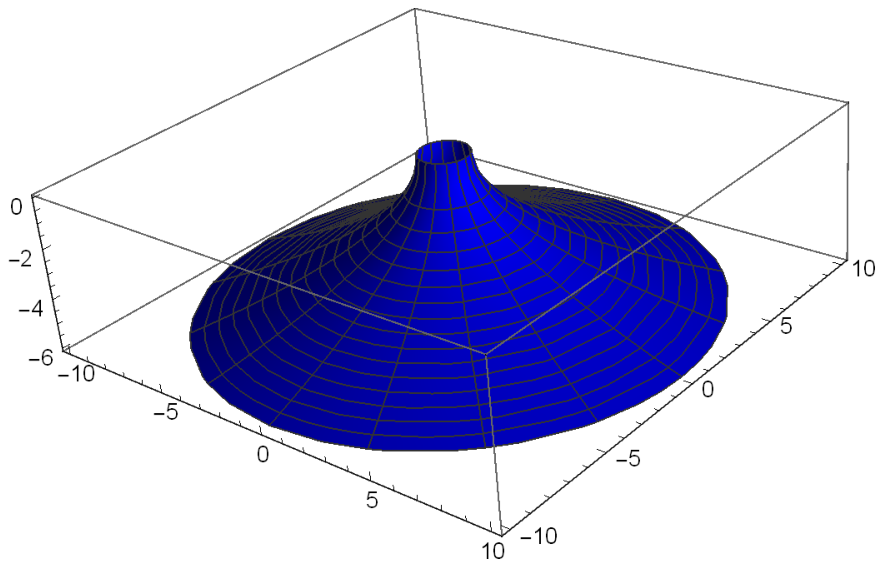,width=0.35\linewidth} &
\end{tabular}
\caption{Plot of embedding diagram for lower universe $h(r)<0$ with respect to radial coordinate with slice $t = const$ and
$\theta = \pi/2$. For a full visualization of the wormhole surface, took a $2\pi$ rotation around the $h-$axis.}\center
\label{Fig:7a4}
\end{figure}
Thus Eqn. ($\ref{713}$) provides asymptotically flat traversable wormholes. We have plotted all the mentioned conditions in Figs. $\ref{Fig:7a1}$ and $\ref{Fig:7a2}$. It can be clearly seen from these figures that our proposed WSF fulfils all the required conditions. To represent the wormhole geometry, we used embedding diagram and excerpt some fruitful information. The spherical symmetry allows us to take an equatorial slice $\theta= \frac{\pi}{2}$ with a rigorous part of time $t=const$. Using these assumptions in Eqn. ($\ref{79}$), we get

\begin{equation}\label{715}
ds^2=\frac{1}{1-\frac{\epsilon(r)}{r}} dr^2+r^2 d\phi^2.
\end{equation}
To visualize this part, embedded Eqn. ($\ref{715}$) into three-dimensional Euclidean space with cylindrical coordinates ($r,~h,~\phi$), given by
\begin{equation}\label{716}
ds^2= dr^2 + dh^2 +r^2 d\phi^2.
\end{equation}
The embedded surface $h\equiv h(r)$ in three-dimensional space with axially symmetry can be written from last equation as
\begin{equation}\label{717}
ds^2= \bigg[1+ \bigg(\frac{dh}{dr} \bigg)^2 \bigg]dr^2 +r^2 d\phi^2.
\end{equation}
By comparing Eqns. ($\ref{716}$) and ($\ref{717}$), one can easily find that
\begin{equation}\label{718}
\frac{dh}{dr}=\pm \bigg( \frac{r}{\epsilon(r) } -1  \bigg)^{\frac{-1}{2}}.
\end{equation}
From Eqn. ($\ref{718}$), we investigate that at the throat the embedded surface is vertical i.e. $\frac{dh}{dr} \rightarrow \infty$. We also examine that away from the throat the space is asymptotically flat because  $\frac{dh}{dr}$ tends to $0$ as $r$ tends to infinity. We have shown the embedded diagram in Figs. $\ref{Fig:7a3}$ and $\ref{Fig:7a4}$.  Moreover, one can visualize the upper universe for $h>0$ and the lower universe $h<0$ in Figs. $\ref{Fig:7a3}$ and $\ref{Fig:7a4}$.

\section{$f(R)$ Gravity}
Our starting point for the $f(R)$ theory of gravity is the Einstein-Hilbert action for modified $f(R)$ gravity,
\begin{equation}\label{1}
\mathcal{S} = \int \sqrt{-g}(\frac{1}{16\pi G}[f(R)]+\mathcal{L}_m) d^4x.
\end{equation}
Here, $\mathcal{L}_m$ stands for the matter Lagrangian field and $f(R)$ is a function of the Ricci scalar $R$. One can derive the field equations by varying the above action with respect to the metric $g_{\zeta \eta}$ as:
\begin{equation}\label{2}
F(R)R_{\zeta \eta}-\frac{1}{2}f(R)g_{\zeta \eta}-\nabla_\zeta \nabla_\eta F(R)+g_{\zeta \eta}\square F(R)=8\pi T^{(m)}_{\zeta \eta},
\end{equation}
where $\square =\nabla_\zeta \nabla^\zeta$ and $\nabla_\zeta$ denotes the covariant derivative. $T^{(m)}_{\zeta \eta}=(\rho, -p_r, -p_t, -p_t)$ represents stress energy momentum tensor, where $\rho$, $p_r$ and $p_t$ are the radial and tangential pressures respectively, and $F(R)=\frac{df}{dR}$. Now by using the trace of energy momentum tensor in above equation and after some manipulations, the above equation can also be written as
\begin{equation}\label{3}
G_{\zeta \eta}=R_{\zeta \eta}-\frac{1}{2}R g_{\zeta \eta}=T^{(eff)}_{\zeta \eta}=\frac{8 \pi}{F} T^{(m)}_{\zeta \eta}+\frac{1}{F}\bigg[\nabla_\zeta \nabla_\eta F(R)-\big(\square F(R)+ \frac{1}{2} R F(R)-\frac{1}{2} f(R)\big)g_{\zeta \eta}\bigg].
\end{equation}
Now using Eqn. ($\ref{3}$) along with eqn. ($\ref{71}$), one can find the following field equations as \cite{bah}:
\begin{eqnarray}
\rho &=& \frac{f}{2} - (1- \frac{\epsilon}{r}) R'^2 f_{RRR} - \frac{f_R}{2 r^2} \bigg[ r \big ( (\epsilon'-4)+ 3 \epsilon \big ) \chi' -2 r (r-\epsilon) \chi'^2  + 2 r (\epsilon-r) \chi'' \bigg] - \frac{f_{RR}}{2 r^2} \bigg[2 r (r - \epsilon) R''
\nonumber\\
&&- (r (\epsilon'-4) +3 \epsilon ) R'   \bigg] \label{d}\\
p_r &=& \frac{-f}{2} + \frac{\epsilon f_R}{2 r^3} \bigg [2 r^2 \chi'^2 + 2 r^2 \chi'' - r \chi' -2  \bigg] +\frac{e^{-2 \chi} f_R}{2 r^2} \bigg[e^{2 \chi} (\epsilon' (r  \chi' +2) -2 r^2 (\chi'^2 + \chi''))   \bigg] + f_{RR} R'
\nonumber\\
&& (1-\frac{\epsilon}{r}) (\chi'+ \frac{2}{r}) ,\label{e}\\
p_t &=& \frac{-f}{2} + \frac{f_R}{2 r^3} \bigg[ \epsilon (2 r \chi' +1) + (\epsilon' - 2 r \chi')r  \bigg] + \frac{f_{RR}}{2 r^2} \bigg[ R' (r (2 r \chi'- \epsilon' +2)- \epsilon (2 r \chi' +1)) +2 r (r- \epsilon) R''    \bigg] +
\nonumber\\
&& f_{RRR} (R'^2 (1- \frac{\epsilon}{r})) . \label{f}
\end{eqnarray}
Here prime denotes the derivative with respect to the radial coordinate $r$. $f_R,~ f_{RR},~ f_{RRR}$ are the first, second and third derivatives of $f(R)$ with respect to the Ricci scalar $R$ respectively. Bronnikov and Starobinsky \cite{bron7} discussed the stability condition for wormhole geometry which is also free from ghosts. It is shown that no realistic wormhole can be formulated in scalar-tensor models for a positive scalar function. In the context of $f(R)$ theory of gravity, the non-existence of wormhole could be violated if $\frac{df}{dR}=F(R)$ is negative \cite{bro}. According to the classical GR, presence of exotic matter in wormhole structure is the main cause of violation of NEC denoted as $\rho+ p_{r} \geq 0,~~   \rho +p_{t} \geq 0$, and WEC denoted as $\rho \geq0,~~ \rho+ p_{r}\geq 0,~~  \rho +p_{t} \geq 0$.  Hochberg and Visser \cite{hoc2,hoc4} extended the results for wormhole solutions, which was earlier described by  Morris-Thorne \cite{mor1}, with exotic matter and provided the result that wormhole throat does not respect NEC. In recent work, our main focus is to calculate the wormhole solutions that respects NEC and also violates the non-existence theorem in the context of $f(R)$ theory of gravity. In view of this approach, we further  simplify our calculation for $\rho+p_r$ and $\rho+p_t$. We found that to respect the NEC

\begin{eqnarray}
\rho + p_r &=& \frac{\epsilon' r- \epsilon}{2 r^3} \bigg ( f_{RR}  R' r + 2 f r \bigg ) ,\label{7e}\\
\rho +p_t &=& f_R \frac{(\epsilon' r+ \epsilon)}{2 r^3} - \frac{ \chi'(\epsilon' r- \epsilon)}{2 r^2}, \label{7f}
\end{eqnarray}
should be positive at the throat. One can notice that first part in Eqn. ($\ref{7e}$) involves flaring out condition which gives $\epsilon' r- \epsilon <0$. Thus 2nd part must be negative for $\rho+ p_r >0$, i.e, $f_{RR}  R' r + 2 f_R <0$. Eqn. ($\ref{7f}$) involves term $f_R= df/dR$, as we consider violation of non-existence theorem (i.e, $df/dR <0$) for spherically symmetric wormhole structure, therefore, for $\rho+ p_t >0$ the remaining term should also be negative. This examination shows the importance of flaring out condition and selection of $f(R)$ gravity model. In next Section, we discuss wormhole solutions under the influence of three different viable $f(R)$ gravity models. It is important to mention that we use $\xi=-1$, $r_0=2$ and $C=1.9$ throughout our further analysis.

\subsection{The Exponential Gravity Model}

%

\begin{figure}\center
\begin{tabular}{cccc}
\epsfig{file=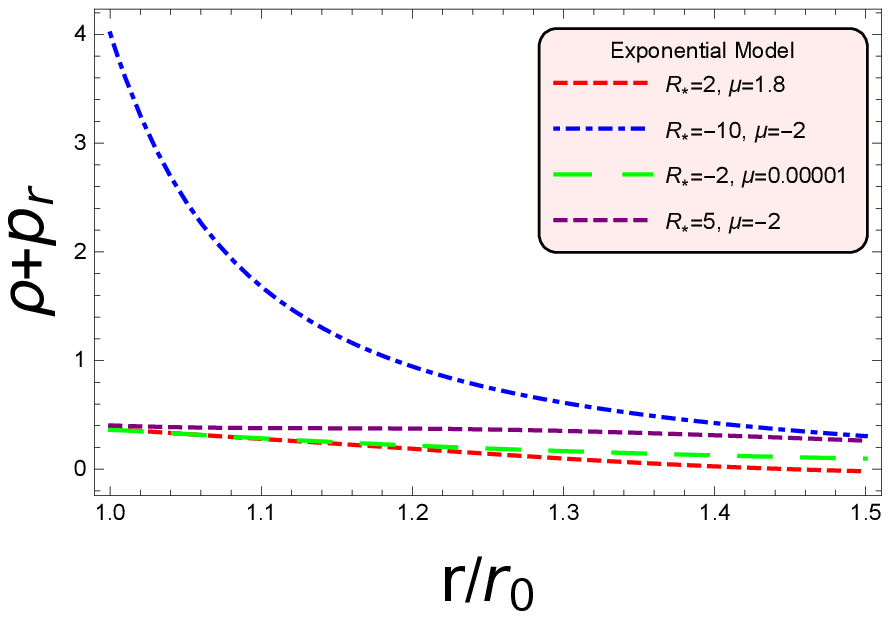,width=0.35\linewidth} &
\epsfig{file=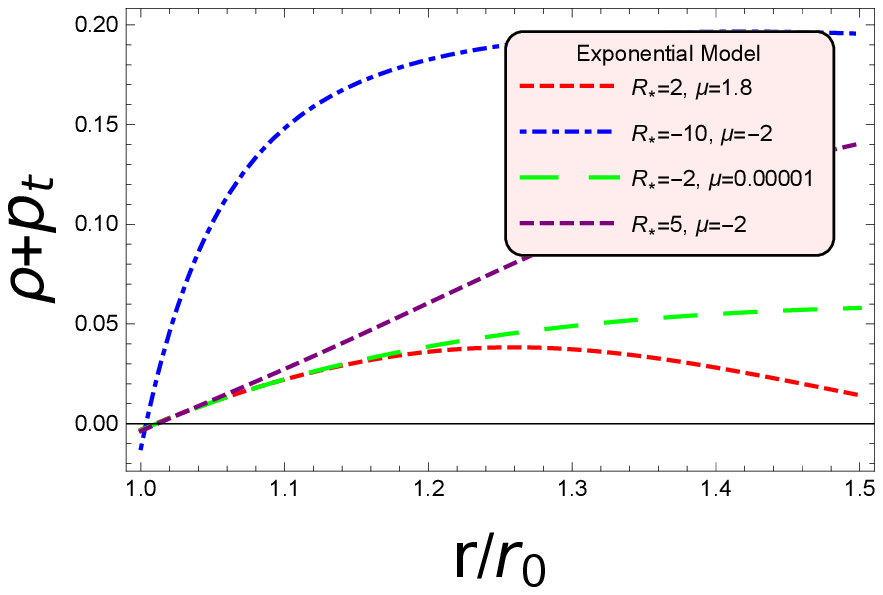,width=0.35\linewidth}\\
\epsfig{file=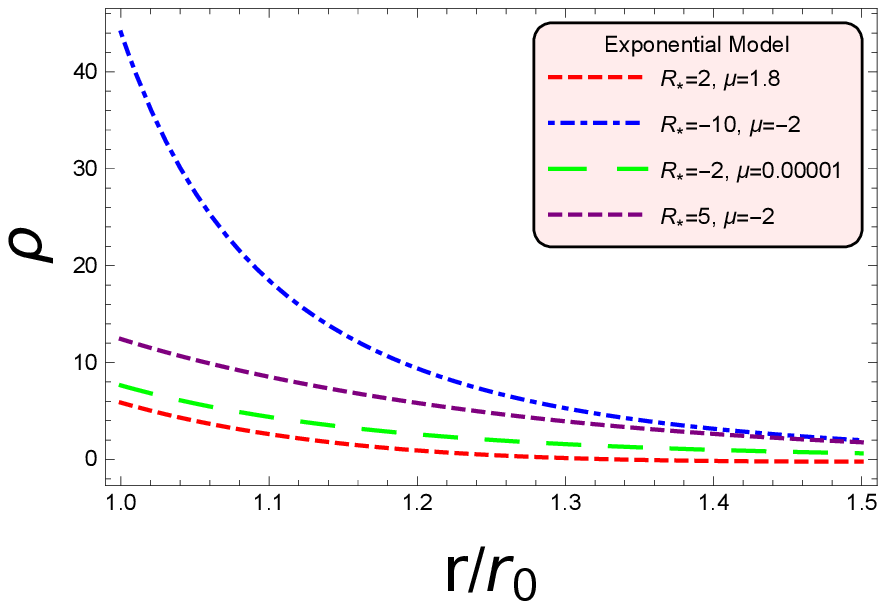,width=0.35\linewidth} &
\epsfig{file=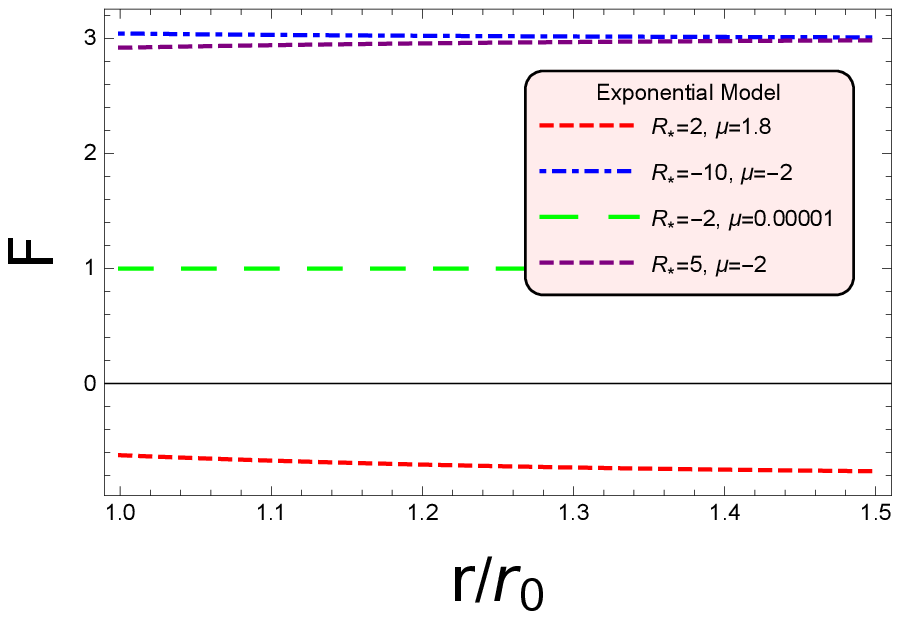,width=0.35\linewidth}\\
\end{tabular}
\caption{The both terms of NEC $\rho + p_r$ (upper left) and $\rho + p_t$ (upper right) are respected at the throat for wormhole geometry for all the considering combinations of free parameters. The first term of WEC $\rho$ (lower left) also shows the validation at the throat. While, we get the violation of non-existence theorem when $\mu>0$, $R_* >0$ (lower right).}\center
\label{Fig:kar1}
\end{figure}

\begin{figure}\center
\begin{tabular}{cccc}
\epsfig{file=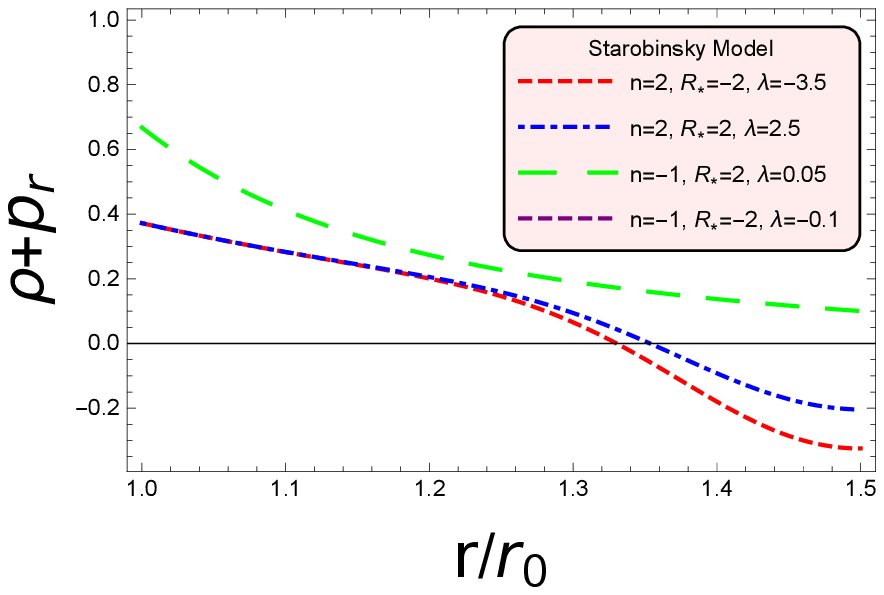,width=0.35\linewidth} &
\epsfig{file=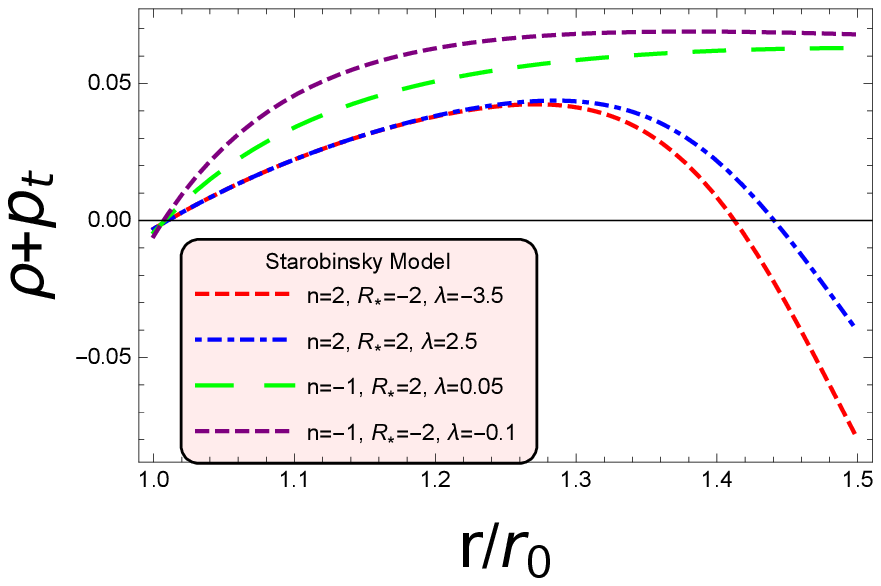,width=0.35\linewidth}\\
\epsfig{file=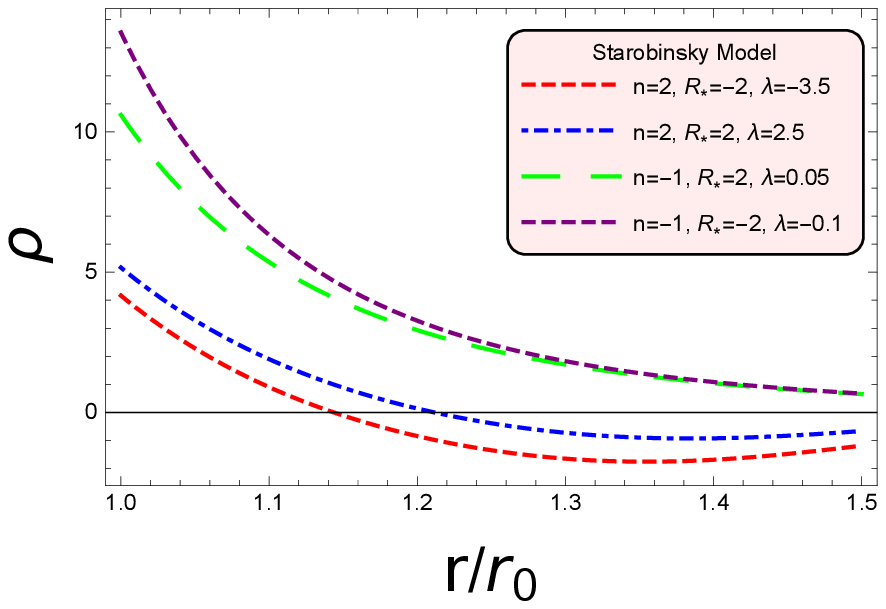,width=0.35\linewidth} &
\epsfig{file=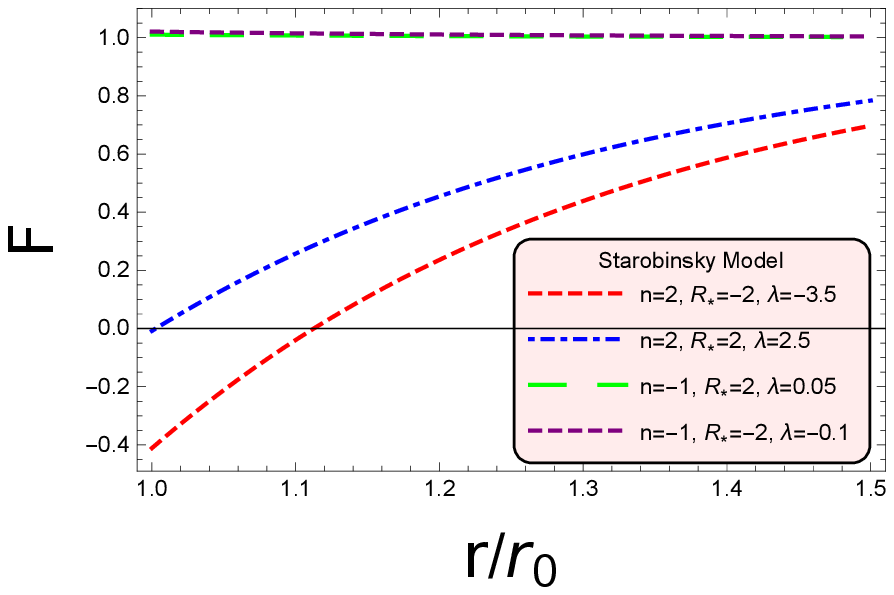,width=0.35\linewidth}\\
\end{tabular}
\caption{$\rho + p_r>0$ (upper left), $\rho + p_t>0$ (upper left) and $\rho>0$ (lower left) for all are considering combinations of $\lambda, R_*$ and $n$. Whereas, $F<0$ for $\lambda<0, R_* <0$ and $n>0$.}\center
\label{Fig:kar2}
\end{figure}

Cognola et al. \cite{cog7} introduced and investigated the exponential gravity model. This model describes the inflation of early universe and accelerated expansion of the current universe in a natural way. The exponential model is defined as

\begin{equation}\label{721}
f(R)=R- \mu R_{*} [1-e^{\frac{-R}{R_*}}],
\end{equation}
where $R_*$ and $\mu$ are arbitrary constants. In particular, we use $\mu=1.8$, $R_* =2$ and evaluate the graphical behaviour of $F=\frac{df}{dR}$, $\rho$, $\rho + p_r$ and $\rho + p_t$. It can be clearly seen from Fig. $\ref{Fig:kar1}$ that the existence of exotic matter can be avoided at the throat for the traversable wormhole geometry. We itemize the analysis below, in the light of exponential $f(R)$ gravity model as:
\begin{itemize}
  \item For $R_*< 0$ and $\mu >0$, WEC is respected for throughout the wormhole geometry, whereas, $F=df/dR >0$ i.e, there is no spherically symmetric wormhole configurations. Such solutions may be interesting for non-static or thin-shell wormholes due to its stable spherically symmetric perturbations \cite{bron7}.
  \item For $R_*< 0$ and $\mu <0$, we found that  $\frac{df}{dR} <0$, whereas, violation of NEC have shown that the wormhole space is filled with exotic matter.
  \item For $R_*> 0$ and $\mu <0$, we get the same result as for $R_*< 0$ and $\mu >0$.
  \item For $R_*> 0$ and $\mu >0$, we found that the existence of exotic matter can be avoided at the throat for the traversable wormhole geometry. It can be clearly seen from Fig. $\ref{Fig:kar1}$, the term $\rho + p_r$ is positive for $1 \leq r/r_0 <1.4$, $\rho + p_t>0$ for $1 \leq r/r_0 <1.7$ and $\rho >0$ for $1 \leq r/r_0 <1.3$. Therefore, NEC is respected for $r/r_0 \in [1,1.4)$ and WEC is respected for $r/r_0 \in [1,1.3)$.
\end{itemize}
Thus, we conclude that $\mu=1.8 $, $R_* =2 $ are the appropriate values to get wormhole solution which violates the non-existence theorem with the presence of negligible amount of exotic matter by using exponential $f(R)$ gravity model with our proposed shape function.

\subsection{Starobinsky $f(R)$ gravity Model}
One of the most well known $f(R)$ gravity model which is consistent
    with cosmological conditions and satisfies solar system and laboratory
    tests, proposed by Starobinsky \cite{star1}, defined as
\begin{equation}\label{719}
f(R)=R + \lambda R_* \big[(1+\frac{R^2}{R_{*}^{2}})^{-n} -1 \big],
\end{equation}
where $\lambda,~ R_*$ and $n$ are free parameters. Model $(\ref{719})$
contains properties of dark energy models and is consistent with
cosmological and local gravity constraints \cite{tsuj}. In investigation, we get the following analysis:
\begin{itemize}
  \item For $n<0$,
  \begin{itemize}
    \item if $R_* <0$ and $\lambda <0$, we get the non-spherically symmetric wormhole geometry with the presence of ordinary matter at the throat i.e., validation of WEC,
    \item we get the same result for $R_* <0$ and $\lambda >0$,
    \item further, we analyze that for $R_* >0$ and $\lambda <0$, again WEC is respected at the throat, whereas, $f>0$.
    \item for $R_* >0$ and $\lambda >0$, we get some interesting results. For $\lambda \in (0,~6)$ have shown an attractive geometry which is feasible for traversable wormhole without exotic matter as $f<0$ with the validation of WEC at the throat. Moreover, when $\lambda >6$, we found that $\rho + p_r >0$, $\rho + p_t >0$ with $\rho<0$.
  \end{itemize}
  \item For $n>0$,
  \begin{itemize}
    \item for the combinations, $R_* <0$, $\lambda <0$ and $R_* >0$, $\lambda <0$, we get the non-existence condition i.e., $F>0$ with $\rho + p_r >0$, $\rho + p_t >0$ and $\rho>0$,
    \item for the combinations, $R_* <0$, $\lambda >0$ and $R_* >0$, $\lambda >0$, we found the violation of non-existence theorem $F>0$. As, WEC is respected at the throat for both the combinations, so, the traversable wormhole inner space is filled with ordinary matter .
  \end{itemize}
\end{itemize}
Thus, we can find the wormhole solutions that may have negligible amount of exotic matter. In particular, we choose $R_*=-2 ,~~ \lambda=-3.5$ and $n=2$ to show the feasible traversable wormhole structure which have an ordinary matter at the throat of the wormhole (see Fig. $\ref{Fig:kar2}$). Further, we have also shown some other combinations which are respected the WEC at the throat with non-asymptotically flat wormhole geometry by using our proposed WSF.

\subsection{Tsujikawa $f(R)$ gravity Model}
\begin{figure}\center
\begin{tabular}{cccc}
\epsfig{file=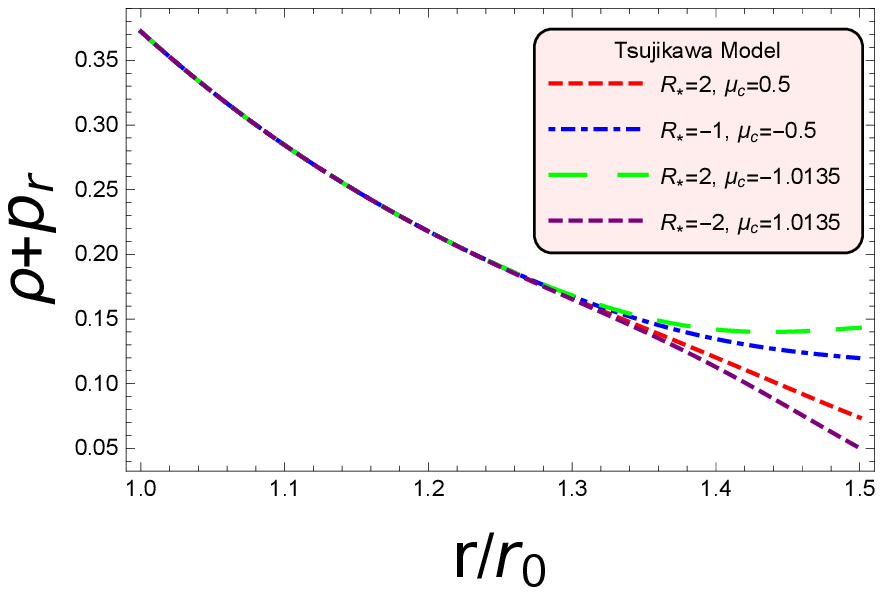,width=0.35\linewidth} &
\epsfig{file=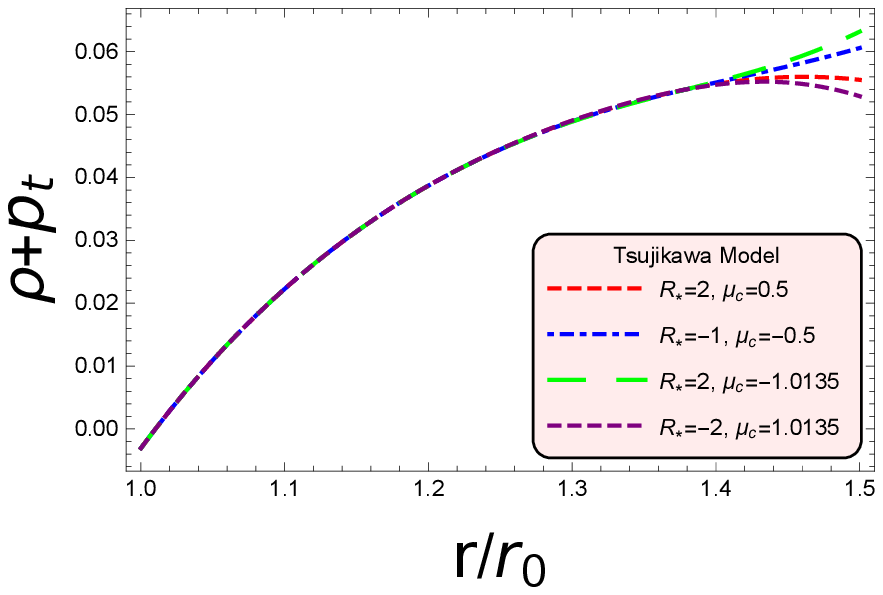,width=0.35\linewidth}\\
\epsfig{file=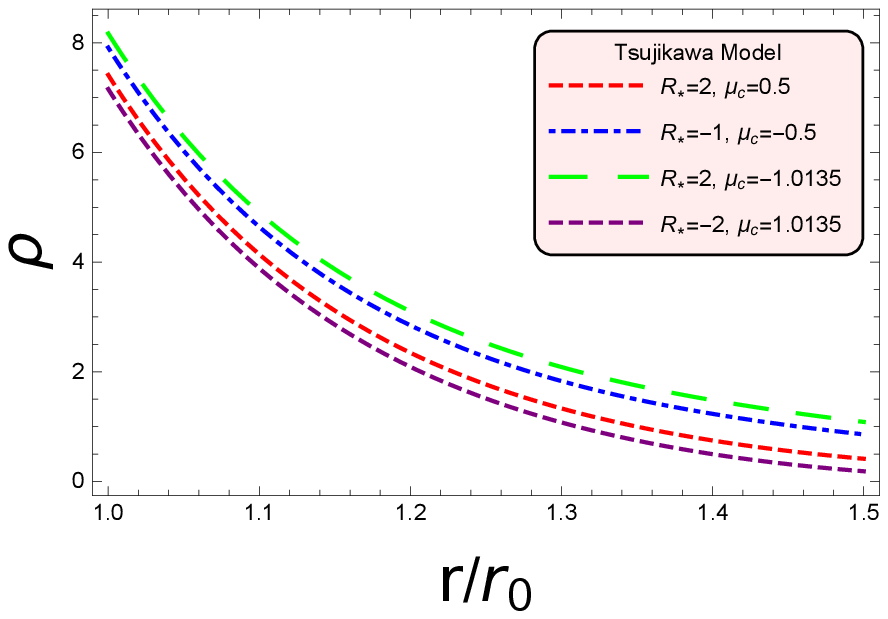,width=0.35\linewidth} &
\epsfig{file=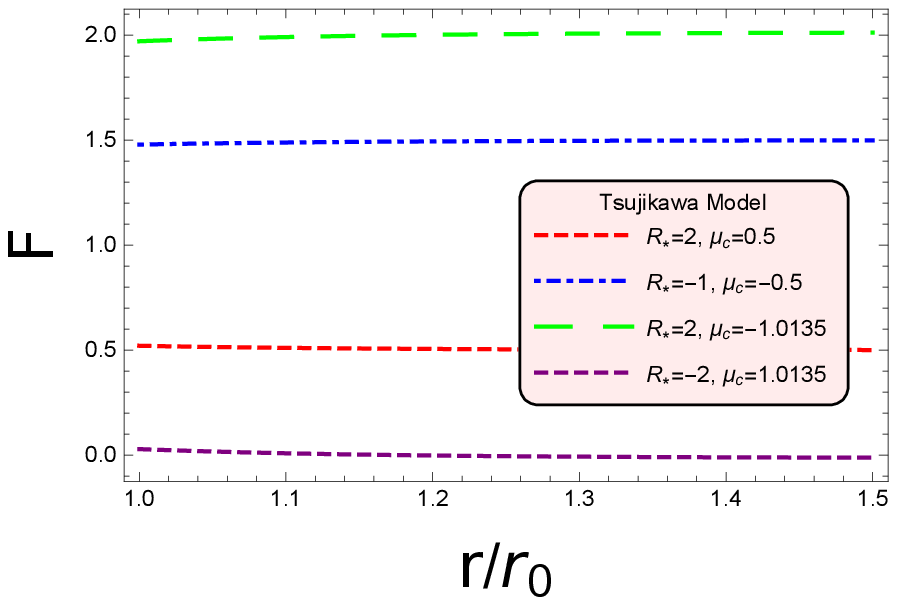,width=0.35\linewidth}\\
\end{tabular}
\caption{The both terms of NEC $\rho + p_r$ (upper left) and $\rho + p_t$ (upper right) are respected at the throat for wormhole geometry for all the considering combinations of free parameters. The first term of WEC $\rho$ (lower left) also shows the validation at the throat. While, we get the violation of non-existence theorem when $R_* <0$ and $\mu_c >0$  (lower right).}\center
\label{Fig:kar3}
\end{figure}

Another viable $f(R)$ model that we consider is known as Tsujikawa model \cite{amen7} and it is defined as

\begin{equation}\label{722}
f(R)=R- \mu_c R_* Tanh (\frac{R}{R_*}).
\end{equation}

Here $R_*$ and $\mu_c $ are arbitrary constants. Tsujikawa \cite{amen7} described that $\mu_c \in (0.905,1)$ to sustain the viability of the model i.e, $ F=\frac{df}{dR} >0$. Whereas for the violation of non-existence theorem of static spherically symmetric wormhole $F=\frac{df}{dR} <0$. To get that requirement we have chosen $\mu_c$ in the neighbourhood of defined range. We assessed the geometric nature of wormhole structure through energy conditions for $\mu_c=1.0135$ and $R_* = -2$. We analyzed the following points during investigation:
\begin{itemize}
  \item For the combinations $R_* <0$, $\mu_c <0$ and $R_* >0$, $\mu_c <0$, WEC is respected for $r/r_0 \in (1,~ \infty)$. This shows that the wormhole inner space is filled with normal matter. Whereas $F>0$ shows the non-spherically symmetric wormhole geometry
  \item For the combinations $R_* <0$, $\mu_c >0$ and $R_* >0$, $\mu_c >0$, we found that $F<0$ and WEC is at the throat. This shows the existence of an attractive wormhole geometry whose vicinity of the throat is filled with ordinary matter.
\end{itemize}
Therefor, it can be seen from the Fig. $\ref{Fig:kar3}$, for the chosen combination of parameters, we get the violation for non-existence theorem i.e, $F<0$. $\rho+ p_r$ and $\rho+ p_t$ both are positive at the wormhole throat therefore NEC is respected which shows the absence of exotic matter at the throat. Moreover, $\rho>0$ also reduces the presence of exotic matter and provides stable and traversable wormhole structure.\\

\section{Concluding Remarks}

In recent years, different researchers proposed WSF that are either ansatz or based on specific theory.
Main purpose of this article is to construct a WSF by employing the KMc. This
condition is mandatory for the spacetime of class one. Several authors extensively consider the KMc to discuss the configurations of spherically symmetric compact objects. In Sec. \textbf{II}, we discuss the construction of WSF in detail. Our proposed WSF fulfils all necessary requirements and provide a viable relativistic configuration for wormhole geometry. We have shown the graphical behaviour of shape funtion and its properties in Figs. $\ref{Fig:7a1}$ and $\ref{Fig:7a2}$. Further, we discuss embedded diagram to represent the wormhole structure. We consider equatorial slice $\theta=\frac{\pi}{2}$, and a fixed moment of time i.e., $t=constant$ for spherical symmetry, and for the visualization we embed it into three dimensional Euclidean space. Moreover, one can visualize the upper universe for $h>0$ and the lower universe $h<0$ in Figs. $\ref{Fig:7a4}$ and $\ref{Fig:7a3}$.


Further, we investigate exact solutions for static spherically symmetric traversable wormhole geometry in the framework of $f(R)$ gravity. For this purpose, we consider four viable $f(R)$ gravity models named as exponential gravity model, Starobinsky gravity Model and Tsujikawa $f(R)$ gravity model to discuss energy conditions and violation of non-existence theorem for wormhole geometry.

\begin{itemize}
\item{Firstly, we consider exponential gravity model which shows the possibility of static spherically symmetric traversable wormhole with violation of non-existence theorem. Fig. $\ref{Fig:kar1}$ have shown that WEC is respected at the throat. Further we investigate that with the variation of $\mu$ either we get the violation of WEC and non-traversable wormhole structure i.e, $df/dr >0 $. Thus we conclude that $\mu>0$, $R_* >0$ is the appropriate combination to get wormhole solution which violates the non-existence theorem with the presence of negligible amount of exotic matter.}
\item{Secondly, we discuss wormhole structure using Starobinsky $f(R)$ gravity model. We discuss different combinations of free parameters in detail and investigate that for the maximum range of free parameters, non-spherically symmetric wormhole geometry is respected the WEC at the vicinity of the throat, such solutions may be interesting for non-static or thin-shell wormholes due to its
    stable spherically symmetric perturbations \cite{bron7}.}

\item{Finally, we consider Tsujikawa gravity model. It can be seen from Fig. $\ref{Fig:kar3}$, for the chosen combination of parameters, we get the violation for nonexistence theorem i.e, $F<0$. $\rho+ p_r$ and $\rho+ p_t$ both are positive at the wormhole throat therefore NEC is respected which shows the absence of exotic matter at the throat. Moreover, $\rho>0$ for larger range of radial coordinate $r$ which also reduce the presence of exotic matter and provides stable and traversable wormhole structure.}

\end{itemize}

Thus, in comparison with Fayyaz and Shamir \cite{fay1}, we have discussed the WSF Eqn ($\ref{713}$) in the context of GR and investigated the presence of exotic matter in the formulation of wormhole. While in the framework of $f(R)$ theory of gravity our results respects NEC and even WEC not at the throat but the larger values of radial coordinate $r$. So, our proposed WSF shows the formation of wormhole geometry with negligible amount of exotic matter.
In future, it would be interesting to check the stability of wormhole solution in other modified gravities for this suggested shape function. As a first step, the study of wormhole geometries using KMc in modified theories with matter coupling is under process.\\\\

\section*{Refrences}



\begin{thebibliography}{70}

\bibitem{ein} Einstein, A. and Rosen, N.: Phys. Rev. \textbf{48}, (1935) 73.\\

\bibitem{fla} Flamm, L.: Comments on Einstein’s theory of gravity, Physikalische Zeitschrift, \textbf{17}, (1916) 448.\\

\bibitem{ell} Ellis, H. G.: Journal of Mathematical Physics, \textbf{14(1)}, (1973) 104-118.\\

\bibitem{whe} Wheeler, J. A.: Phys. Rev. \textbf{97}, (1955) 511.\\

\bibitem{kar7} Kar, S.: Phys. Rev. D \textbf{49(2)}, (1994) 862.\\

\bibitem{kar72} Kar, S., and Sahdev, D.: Phys. Rev. D \textbf{53(2)}, (1996) 722.\\

\bibitem{vac} Vacaru, S, I., Douglas, S., Vitalie, A. B., and Denis A. D.: Phys. Lett. B \textbf{519}, (2001) 249.\\

\bibitem{Dzh} Dzhunushaliev, V., Vladimir, F., Douglas, S. and Ratbay, M.: Phys. Rev. D \textbf{82}, (2010) 045032.\\

\bibitem{Vis} Visser, M., Kar, S., and Naresh, D.: Phys. Rev. Lett. \textbf{90}, (2003) 201102.\\

\bibitem{can7} Canfora, F.,  Dimakis, N. and Paliathanasis, A.: Phys. Rev. D \textbf{96}, (2017) 025021.\\

\bibitem{mor1} Morris, M. S. and Thorne, K. S.: Am. J. Phys. \textbf{56}, (1988) 395.\\

\bibitem{mor2}  Morris, M. S. and Thorne, K. S. and Yurtsever, U.: Phys. Rev. Lett. \textbf{61}, (1988) 1446.\\

%
%

\bibitem{nan}   Nandi,K. K.,  Islam, A. and Evans,  J.: Phys. Rev. D \textbf{55}, (1997) 2497.\\

\bibitem{lob}  Lobo, F. S. N. and Oliveira, M. A.: Phys. Rev. D \textbf{81}, (2010) 067501.\\

\bibitem{ric} Richarte,  M. and Simeone, C.: Phys. Rev. D \textbf{80}, (2009) 104033.\\

\bibitem{meh1} Mehdizadeh, M. R., Zangeneh, M. K. and Lobo, F. S. N.: Phys. Rev. D \textbf{91}, (2015) 084004.\\

\bibitem{dzh}Dzhunushaliev, V.D.  and Singleton, D.: Phys. Rev. D \textbf{59}, (1999) 064018.\\

\bibitem{sha} Shaikh, R., and Kar, S.: Phys. Rev. D \textbf{94}, (2016) 024011.\\

\bibitem{meh2} Mehdizadeh, M. R. and Ziaie, A.H.: Phys. Rev. D \textbf{95}, (2017) 064049.\\

\bibitem{Buch} Buchdahl, H. A.: Mon. Not. R. Astron. Soc. \textbf{150}, (1970) 1.\\

\bibitem{cap1} Capozziello, S. and Francaviglia, M.: Gen. Relt. Grav. \textbf{40}, (2008) 357.\\

\bibitem{cap2} Capozziello, S., Carloni, S. and Troisi, A.: Recent Res. Dev. Astron. Astrophys.  \textbf{1}, (2003) 625.\\

\bibitem{har} Harko, T., Lobo, F. S. N., Mak, M. K. and Sushkov S. V.: Phys. Rev. D  \textbf{87}, (2013) 067504.\\

\bibitem{rah} Rahaman, F., Banerjee, A., Jamil, M., Yadav, A. K. and Idris, H., Int. J. Theor. Phys.  \textbf{53}, (2014) 1910.\\

\bibitem{jam} Jamil, M., Rahaman, F., Myrzakulov, R., Kuhfittig, P.K.F. and Ahmed, N., Mondal, U.F.:  J. Korean Phys. Soc.
\textbf{65}, (2014) 917.\\




\bibitem{bron7} Bronnikov, K. A. and Starobinsky, A. A.:  JETP letters, \textbf{85(1)}, (2007) 1.\\

\bibitem{bro} Bronnikov, K. A., Skvortsova, M. V., and Starobinsky, A. A.: Gravitation and Cosmology, \textbf{16(3)} (2010) 216-222.\\

\bibitem{bah} Bahamonde, S., Jamil, M., Pavlovic, P. and Sossich,  M.: Phys. Rev. D \textbf{94}, (2016) 044041.\\

\bibitem{goda} Godani, N. and Samanta, G. C.: Int. J. Mod. Phys. D \textbf{28}, (2018) 1950039.\\

\bibitem{jah} Jahromi, A. S.  and Moradpour, H.: Int. J. Mod. Phys. D \textbf{27}, (2018) 1850024.\\

\bibitem{cata} Cataldo, M., Liempi, L. and Rodr´ıguez, P.: Phys. Lett. B \textbf{757}, (2016) 130-135.\\

\bibitem{sam} Samanta, G. C., Godani, N. and Bamba, K.: Int. J. Mod. Phys. D \textbf{29(09)}, (2020) 2050068.\\

\bibitem{jam2} Jamil, M., Momeni, D., Myrzakulov, R.: Eur. Phys. J. C \textbf{73}, (2013) 2267.\\

\bibitem{shar2} Sharif, M. and Zahra, Z.: Astrophysics and Space Science, \textbf{348(1)}, (2013) 275.\\

\bibitem{sham3} Shamir, M. F. and Zia, S.: Astrophysics and Space Science \textbf{363}, (2017) 247.\\

\bibitem{gol} Golchin, H. and Mehdizadeh, M. R.: Eur. Phys. J. C,  \textbf{79(9)}, (2019) 777.\\

\bibitem{karm} Karmarkar, K. R.: Proc. Indian Acad. Sci. A \textbf{27} (1948) 56.\\

\bibitem{6abb} Abbas, G., Qaisar, S., Javed, W., and Meraj, M. A.:  IRAN J SCI TECHNOL A \textbf{42(3)}, (2018) 1659.\\

\bibitem{6ful} Fuloria, P., and Pant, N.:  Eur. Phys. J. A \textbf{53(11)}, (2017) 227.\\

\bibitem{6kuh} Kuhfittig, P. K.: Annals of Physics, \textbf{392}, (2018) 63-70.\\

\bibitem{6bah} Bhar, P., Singh, K. N., and Manna, T.: Int. J. Mod. Phys. D \textbf{26(09)}, (2017) 1750090.\\

\bibitem{6ged} Gedela, S., Bisht, R. K., and Pant, N.: Eur. Phys. J. A \textbf{54(11)}, (2018) 207.\\

\bibitem{kuhf} Kuhfittig, P. K. F.: Pramana \textbf{92} (2019) 75.\\

\bibitem{fay1} Fayyaz, I., Shamir, M. F.: Chin. J. Phys. \textbf{66}, (2020) 553.\\


\bibitem{anch} Anchordoqui, L.A., Torres, D.F., Trobo, M.L., Bergliaffa, S.E.P.: Phys. Rev. D \textbf{57},  (1998) 829.\\


\bibitem{hoc2} Hochberg, D. and Visser, M.: Phys. Rev. \textbf{D58}, (1998b) 044021 .\\

\bibitem{hoc4} Hochberg, D. and Visser, M: Phys. Rev. Lett. \textbf{81}, (1998a) 746.\\

\bibitem{cog7} Cognola, G., Elizalde, E., Nojiri, S., Odintsov, S. D., Sebastiani, L., and Zerbini, S.: Phys. Rev. D  \textbf{77(4)}, (2008) 046009.\\

\bibitem{star1} Starobinsky, A. A.,
 JETP letters, \textbf{86(3)}, (2007) 157.\\


 \bibitem{tsuj} Tsujikawa, S.,
    In Lectures on Cosmology, Springer, Berlin, Heidelberg, 2010, 99-145.
 \\

\bibitem{amen7} S. Tsujikawa, Phys. Rev. D. \textbf{77}, (2008) 023507. \\
\end{thebibliography}
\end{document}